\documentclass[fleqn,10pt]{wlscirep}
\usepackage[utf8]{inputenc}
\usepackage[T1]{fontenc}
\title{Circulator function  in a Josephson junction circuit and  braiding of Majorana zero modes}

\author{Mun Dae Kim}
\affil{College of Liberal Arts, Hongik University, Sejong 30016, Korea}
\affil{mundkim@gmail.com}

\begin{abstract}
We propose a scheme for the circulator function in a superconducting circuit
consisting of a three-Josephson junction loop and a trijunction.
In this study we obtain the exact Lagrangian of the system by deriving
the effective potential from the fundamental boundary conditions.
We subsequently show that we can selectively choose the direction of current
flowing through the branches connected at the trijunction,
which performs a circulator function.
Further, we use this circulator function for  a non-Abelian
braiding of Majorana zero modes (MZMs). In the branches of the system
we introduce pairs of MZMs  which interact with each other through
the phases of trijunction. The circulator function determines the
phases of the trijunction and thus the coupling between the MZMs
to gives rise to the braiding operation. We modify the system so
that MZMs  might be coupled to the external ones to perform
qubit operations in a scalable design.
\end{abstract}

\begin{document}

\flushbottom \maketitle
%
%
\thispagestyle{empty}



While the ultimate goal of  practical quantum computer is still
far away, the noisy intermediate-scale quantum (NISQ) computing
\cite{Preskill} is expected to be realized in the near future due
to the remarkable advancement in the qubit coherence and control.
The quantum supremacy that  quantum device can solve a problem
that no classical computer can solve in any feasible amount of
time is regarded as a notable milestone. \cite{Arute}
%
The programmable NISQ computing for  quantum supremacy requires
a scalable design of quantum circuit,  which is severely challenging.
We, here, provide an approach to cope with this challenge by proposing a scheme for a
circulator function which enables  selective coupling between
arbitrary two branches at a trijunction by using a
three-Josephson junction flux qubit as a control element in a
superconducting circuit. \cite{Schmidt,Underwood,Koch,Nunnenkamp}

In this study we introduce a three-Josephson junction loop
consisting of three small loops with three branches and a
trijunction as shown in Fig. \ref{scheme}(a). Usually the
Hamiltonians of the superconducting circuit with threading fluxes
for quantum information processing have been provided
phenomenologically. The effective potential in the Hamiltonian is
given in an approximate way so that the form and the coefficients
have not been precisely derived from the first principle. For the
understanding of the system we need to know the exact form of  the
Hamiltonian and the process by which the Hamiltonian is obtained.
For the superconducting loop system in the present study we
derive the Lagrangian of system exactly from fundamental boundary
conditions and obtain the effective potential of the system
analytically. This Lagrangian describes the circulating function
in the ground state of the system, where we can selectively couple
two branches to flow currents while the other branch does not.
This kind of study will help analyzing other systems for
quantum information processing.

Circulator is a nonreciprocal three-port device that routes a
signal to the next port. For the universal quantum computing
quantum gates between different two qubits in a scalable design
is required. Hence the circulator function which enables selective coupling between arbitrary two qubits
among several qubits has been studied intensively.
Recently Josephson junction based on-chip circulators much smaller
than commercial microwave circulator have been proposed for the
quantum information processing with superconducting devices.
\cite{Koch,Sliwa} The superconductor-based circulators have
remarkably small photon losses compared to the commercial
nonreciprocal ferrite circulators \cite{Pozar}. Moreover, the
superconductor-based circulators are much smaller than the
commercial circulators so that they can be integrated into a
scalable circuit.

By piercing a magnetic flux into one of three small loops we are
able to make the current  flow between two branches selectively
{\it in situ}, while the other is isolated, resulting in  the
circulator function. Usually the circulator routes a signal from
one port to the other. Present design, in contrast,  performs a
circulator function that routes a signal between two branches at a
trijunction in a closed circuit rather than transferring the
signal  to outer port.  In this way, we can connect arbitrary pair
of branches to perform quantum gate operations.
For the NISQ computing we need to perform the circulator function  in a scalable circuit
where the trijunctions are connected with each other to form a lattice structure.
We thus consider an improved design  where the trijunction is located outside of the loop as shown in Fig.
\ref{scheme}(b), which is topologically equivalent with the design in Fig. \ref{scheme}(a).

Further, we can use the circulator function to realize the braiding of
Majorana zero modes (MZM) \cite{Aasen,Beenakker} for topological
quantum computing. \cite{Nayak,Lahtinen}
Topologically-protected quantum processing is expected to provide
a path towards fault-tolerant quantum computing. Since quantum
states are susceptible to environmental decoherence, protection
from local perturbation is an emergent challenge for quantum
information processing. Non-Abelian states are the building block
of topological quantum computing carrying the nonlocal
information. The nonlocally encoded quantum information is
resilient to local noises and, if the temperature is smaller than
the excitation gap, temporal excitation rate is exponentially
suppressed.
Majorana zero modes, $\gamma$, are  predicted to exhibit
non-Abelian exchange statistics, and they are  self-adjoint
$\gamma^\dagger=\gamma$ in contrast to ordinary fermion operators.
The theoretically proposed structures attracted a great deal of
intention to realizing MZMs in condensed matter systems. MZMs are
predicted to emerge in $\nu=5/2$ fractional quantum Hall states,
\cite{Nayak,Sarma} p-wave superconductors,
\cite{Stanescu11,Stanescu12} and one- or two-dimensional
semiconductor/superconductor hybrid structures.
\cite{Lutchyn} The branches in our scheme
for braiding contains semiconductor/superconductor hybrid structures
with  p-wave-like superconductivity  induced  from s-wave
superconductors via proximity effect.

In two-dimensional spinless $p+ip$ topological superconductors MZMs are hosted in vortices or in the chiral  edge modes
as  localized Andreev-bound zero-energy states at the Fermi energy.
The p-wave-like superconductivity  can be induced  from s-wave
superconductors via proximity effect in a hybrid structure.
\cite{Fu} Semiconductor thin film with  Zeeman splitting and
proximity-induced s-wave superconductivity has been expected to be
a  suitable platform for hosting MZMs. \cite{Bocquillon}
On the other hand, the one-dimensional semiconducting nanowire  has also been shown to
provide MZMs at the ends of the nanowire. \cite{Zhang}
The MZMs should be prepared, braided, and fused to implement qubit operations. In
one-dimensional wire the braiding is not well defined, which can
be overcome in a wire network of trijunction. However, the
original scheme \cite{Fu}  with Josephson trijunction has not yet
been explored.

Recently, an experimental evidence of MZM in a trijunction has been reported. \cite{Yang}
The nanowire trijunctions are manipulated by the chemical potential,  \cite{Harper}
the charging energy, \cite{Heck} and the phase.   \cite{Stenger}
In the present study a pair of MZMs can be introduced in each branch near the trijunction of Fig.
\ref{scheme}(b). Three MZMs of each pair are coupled through  Josephson junctions with
phase differences $\varphi'_1,\varphi'_2,$ and $\varphi'_3$ in the system.
The three Josephson junction loop controls the selective coupling among three MZM pairs.
By applying a threading flux into one of the loops Fig. \ref{scheme}(b)
we can use the circulator function to control the phases $\phi'_i$ and thus
the couplings among MZMs in the trijunction
to perform the braiding operation and, further, quantum gate operations.
In contrast to the previous phase modulation scheme \cite{Stenger}
trying to switch off the current mediated by MZMs which are inside
of the loop the present proposal uses circulating function to
perform braiding operations. Further, our scheme enables the
interaction between MZMs outside so that we may provide a scalable
design in a one or two-dimensional lattice system for coupling
between MZMs which belong to different trijunctions.




\section*{Results}

\subsection*{Three-Josephson junction loop with a trijunction.}


The precise fluxoid quantization condition of superconducting loop
reads $-\Phi_t+(m_c/q_c)\oint{\vec v}_c\cdot d{\vec l}=n\Phi_0$
with ${\vec v}_c$ being the average velocity of Cooper pairs,
$q_c=2e$ the Cooper pair charge, and $m_c=2m_e$ the Cooper pair
mass. \cite{Tinkham,Kim04} The total magnetic flux $\Phi_t$ threading the loop is the
sum of the external and the induced flux $\Phi_t=\Phi_{\rm
ext}+\Phi_{\rm ind}$. With the superconducting unit flux quantum
$\Phi_0=h/2e$ we introduce the reduced fluxes,
$f_t=\Phi_t/\Phi_0=f+f_{\rm ind}$ with $f=\Phi_{\rm ext}/\Phi_0$
and $f_{\rm ind}=\Phi_{\rm ind}/\Phi_0$, expressing the fluxoid
quantization condition as $kl=2\pi(n+f_t)$ with $l$ being the
circumference of  the loop, $k$ the wave vector of the Cooper
pair wavefunction and $n$ an integer.

\begin{figure}[t]
\vspace{-2cm}
\hspace{-1cm}
\includegraphics[width=1.2\linewidth]{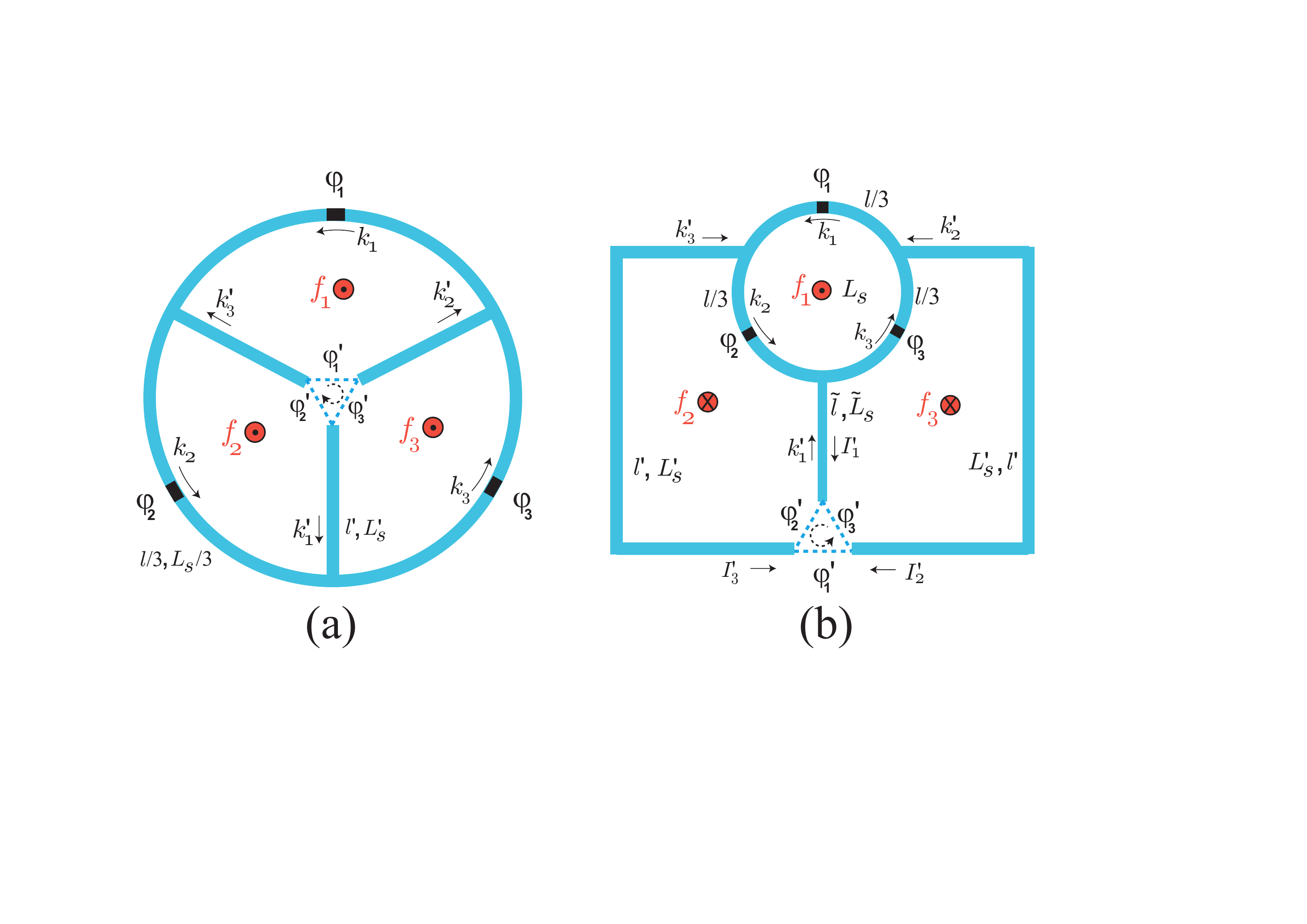}
\vspace{-4cm} \caption{(a)Three-Josephson junction loop with
length $l$ and geometric inductance $L_s$  has three Josephson
junctions with  phase differences $\varphi_i$ and  three
branches with length $l'$ and geometric inductance $L'_s$. $k_i$ and
$k'_i$ are the wave vectors of the Cooper pairs and $f_i$  the
external flux threading the loops. $\varphi'_i$'s are  the
trijunction phase differences. (b) A scheme that three branches and trijunction
are extracted out from the three-Josephson junction loop and turned over:  left and
right branches have length $l'$ and geometric inductance $L'_s$ and
central branch  ${\tilde l}$ and ${\tilde L}_s$. Two
schemes in (a) and (b) are topologically equivalent with each
other. } \label{scheme}
\end{figure}

The scheme in Fig. \ref{scheme}(a) consists of three-Josephson junction loop
and  three small loops with threading fluxes  $f_i=\Phi_{\rm ext,i}/\Phi_0$.
The fluxoid quantization conditions around three loops,  including the phase differences $\varphi_i$ and $\varphi'_i$
across the Josephson junctions,  are represented as the following periodic boundary conditions, \cite{Kim11,Kim17}
\begin{eqnarray}
\label{center}
k_1\frac{l}{3}-k'_3l'+k'_2 l'+\varphi_1+\varphi'_1&=&2\pi(m_1+f_1+f_{\rm ind,1}),\\
\label{right}
k_2\frac{l}{3}-k'_1l'+k'_3 l'+\varphi_2+\varphi'_2&=&2\pi(m_2+f_2+f_{\rm ind,2}),\\
\label{left}
k_3\frac{l}{3}-k'_2 l'+k'_1l'+\varphi_3+\varphi'_3&=&2\pi(m_3+f_3+f_{\rm ind,3}),
\end{eqnarray}
where $k_i, l$, and $l'$ are  the wave vector of Cooper pairs, the
length of  the three-Josephson junction loop, and three
branches, respectively, and $m_i$'s are integer. Here,
$\varphi_i$'s are the phase differences of Josephson junctions in
the three-Josephson junction loop and
$\varphi'_i$'s phase differences of the  trijunction whose
positive direction, we choose, is clockwise as shown in Fig.
\ref{scheme}(a). Which branches carry current, while  the other
not, is determined by  threading a flux, $f_i$, into a specific loop.

The induced flux  $f_{\rm ind,1}$, for example, can be written as
$f_{\rm ind,1}=\Phi_{\rm ind,1}/\Phi_0=(1/\Phi_0)(L_sI_1/3+L'_sI'_2-L'_sI'_3)$,
where the Cooper pair current $I_i$ is given by
\begin{eqnarray}
\label{Ik}
I_i=-(n_cAq_c/m_c)\hslash k_i
\end{eqnarray}
with the Cooper pair density $n_c$ and the cross section $A$
of the loop. The induced flux, $\Phi_{\rm ind,1}$, consists of contributions from three conducting lines,
$L'_sI'_3, L'_sI'_2$ and $L_sI_1/3$, where  $L_s$ and $L'_s$ are the geometric inductance
of the three-Josephson junction loop and a branch, respectively, and the inductance of one third of the loop contributes to the induced flux.
Further, we introduce the kinetic inductances $L_K=m_cl/An_cq^2_c$ and $L'_K=m_cl'/An_cq^2_c$, \cite{Kim04,Meservey,Hazard}
and then the induced fluxes  become
$f_{\rm ind,1}=-(1/2\pi)[(L'_s/L'_K)(k'_2-k'_3)l'+(L_s/L_K)k_1l/3]$,
$f_{\rm ind,2}=-(1/2\pi)[(L'_s/L'_K)(k'_3-k'_1)l'+(L_s/L_K)k_2l/3]$,
and $f_{\rm ind,3}=-(1/2\pi)[(L'_s/L'_K)(k'_1-k'_2)l'+(L_s/L_K)k_3l/3]$
to represent the boundary conditions as
\begin{eqnarray}
\label{cbc}
\left(1+\frac{L_s}{L_K}\right)k_1\frac{l}{3}+\left(1+\frac{L'_s}{L'_K}\right)(k'_2-k'_3)l'
&=&2\pi\left(m_1+f_1-\frac{\varphi_1+\varphi'_1}{2\pi}\right)\\
\label{rbc}
\left(1+\frac{L_s}{L_K}\right)k_2\frac{l}{3}
+\left(1+\frac{L'_s}{L'_K}\right)(k'_3-k'_1)l'&=&2\pi\left(m_2+f_2-\frac{\varphi_2+\varphi'_2}{2\pi}\right)\\
\label{lbc}
\left(1+\frac{L_s}{L_K}\right)k_3\frac{l}{3}+\left(1+\frac{L'_s}{L'_K}\right)(k'_1-k'_2)l'
&=&2\pi\left(m_3+f_3-\frac{\varphi_3+\varphi'_3}{2\pi}\right).
\end{eqnarray}

In the system of Fig. \ref{scheme}(a) three Josephson junctions with $\varphi'_i$ compose a trijunction
which satisfies the periodic boundary condition $\varphi'_1+\varphi'_2+\varphi'_3=2\pi n'$ with an integer $n'$.
By using this condition and summing above three equations we can check that the boundary condition for three-Josephson junction loop
can be expressed as
$\left(1+L_s/L_K\right)(k_1+k_2+k_3)(l/3)=2\pi\left[n+f_1+f_2+f_3-(\varphi_1+\varphi_2+\varphi_3)/2\pi\right]$
with an integer $n$, which can also be derived directly from the fluxoid quantization condition.
If we assume the superconducting branches in Fig. \ref{scheme}(a) have  the same cross section $A$ and
Cooper pair density $n_c$ in Eq. (\ref{Ik}), the current
conservation conditions, $I_1=I_3+I'_2, I_2=I_1+I'_3$, and
$I_3=I_2+I'_1$,  at the nodes of three-Josephson junction loop
give rise to the relations,
\begin{eqnarray}
\label{kcond}
k_1=k_3+k'_2, ~~k_2=k_1+k'_3, ~~k_3=k_2+k'_1.
\end{eqnarray}
From the boundary conditions in Eqs. (\ref{cbc})-(\ref{lbc}) in
conjunction with the relations in Eq. (\ref{kcond}) we can readily
obtain $k_i$ and $k'_i$ in terms of $\varphi_i$ and
$\varphi'_i$ as
\begin{eqnarray}
\label{ki}
k_i&=&\frac{2\pi}{l}\frac{3L_K}{L'_{\rm eff}}\left(m_i+f_i-\frac{\varphi_i+\varphi'_i}{2\pi}\right)
+\frac{2\pi}{l}\left(\frac{L_K}{L_{\rm eff}}-\frac{L_K}{L'_{\rm eff}}\right)\left(n+f_1+f_2+f_3-\frac{\varphi_1+\varphi_2+\varphi_3}{2\pi}\right),\\
\label{k'i}
k'_i&=&\frac{2\pi}{l}\frac{3L_K}{L'_{\rm eff}}\left(m_{i+2}-m_{i+1}+f_{i+2}-f_{i+1}-\frac{\varphi_{i+2}+\varphi'_{i+2}}{2\pi}+\frac{\varphi_{i+1}+\varphi'_{i+1}}{2\pi}\right),
\end{eqnarray}
where the effective inductances are defined as $L_{\rm eff}\equiv L_K+L_s$ and $L'_{\rm eff}\equiv L_K+L_s+9(L'_K+L'_s)$.
Here and after,  the indices, $i$, are modulo 3, for example, $i+1=i+1~ {\rm mod} ~3$.

The dynamics of Josephson junction is described by the
capacitively-shunted model, where  the current relation is given
by $I=-I_c\sin\phi+C{\dot V}=-I_c\sin\phi-C(\Phi_0/2\pi){\ddot
\phi}$ with  the critical current $I_c$, the capacitance $C$ of Josephson
junction, and the voltage-phase relation,
$V=-(\Phi_0/2\pi){\dot \phi}$. The quantum Kirchhoff relation then
becomes
$-(\Phi^2_0/2\pi L_K)(l/2\pi)k_i=-E_{J}\sin\phi_i-C(\Phi_0/2\pi)^2{\ddot \phi}_i$
with the Josephson coupling energy  $E_J=\Phi_0I_c/2\pi$  and
the current  $I=-(n_cAq_c/m_c)\hslash k$.
From the Lagrangian ${\cal L}=\sum_i(1/2)C_i(\Phi_0/2\pi)^2{\dot \phi}^2_i-U_{\rm eff}(\{\phi_i\})$
with the effective potential of the system, $U_{\rm eff}(\{\phi_i\})$,
the equation of motion, $C_i(\Phi_0/2\pi)^2{\ddot\phi}_i=-\partial U_{\rm eff}/\partial\phi_i$,
can be derived from the Lagrange equation $(d/dt)\partial {\cal L}/\partial {\dot \phi}_i-\partial{\cal L}/\partial\phi_i=0$.
By using the quantum Kirchhoff relation  the equation of motion then can be represented as
\begin{eqnarray}
\label{QK}
\frac{\Phi^2_0}{2\pi L_K}\frac{l}{2\pi}k_i-E_{J}\sin\phi_i=-\frac{\partial U_{\rm eff}}{\partial\phi_i}.
\end{eqnarray}

We  can construct the effective potential $U_{\rm eff}(\{\varphi_i,\varphi'_i\})$ as follows,
\begin{eqnarray}
\label{Ufin}
U_{\rm eff}(\{\varphi_i,\varphi'_i\})&=&\frac{3\Phi^2_0}{2L'_{\rm eff}}\left[\left(m_1+f_1-\frac{\varphi_1+\varphi'_1}{2\pi}\right)^2+
\left( m_2+f_2-\frac{\varphi_2+\varphi'_2}{2\pi}\right)^2+\left(m_3+f_3-\frac{\varphi_3+\varphi'_3}{2\pi}\right)^2   \right]\nonumber\\
&+&\left(\frac{\Phi^2_0}{2L_{\rm eff}}-\frac{\Phi^2_0}{2L'_{\rm eff}}\right)\left(n+f_1+f_2+f_3-\frac{\varphi_1+\varphi_2+\varphi_3}{2\pi}\right)^2-\sum^3_{i=1}(E_{J}\cos\varphi_i+E'_{J}\cos\varphi'_i),
\end{eqnarray}
which consists of the inductive energies of the loops and
Josephson junction energies with $E'_J$ being the Josephson
junction energy of trijunction.
We can easily check that the effective potential $U_{\rm eff}(\{\varphi_i,\varphi'_i\})$
satisfy  the equation of motion in Eq. (\ref{QK}) for $\phi_i=\varphi_i$ with $k_i$'s in Eq. (\ref{ki}).
The kinetic inductance $L_K$ is much smaller than the geometric inductance $L_s$.
For the usual parameter regime for three-Josephson junction qubit $L_K/L_s\sim O(10^{-3})$ \cite{Wal}
so that we can approximate the effective inductances as $L_{\rm eff}\approx L_s$ and $L'_{\rm eff}\approx L_s+9L'_s$.

Further, the effective potential $U_{\rm
eff}(\{\varphi_i,\varphi'_i\})$ should also satisfy the quantum
Kirchhoff relation for the phase variables $\varphi'_i$. In Fig.
\ref{scheme}(a) we consider the currents ${\tilde I}_i$ across the
Josephson junction with phases $\varphi'_i$ and $I'_i$ flowing in
the branch, where  the direction of $\tilde{I}_i$ is  counterclockwise and $I'_i$ is opposite to $k'_i$ (See
Fig. S1(a) in the Supplementary Information). Then with the current conservation relation at
nodes, $I'_i={\tilde I}_{i+2}-{\tilde I}_{i+1}$, and the current
relation of Josephson junction, ${\tilde
I}_i=-I'_c\sin\varphi'_i-C'(\Phi_0/2\pi){\ddot \varphi'_i}$, we
have   $I'_i=-(I'_c\sin\varphi'_{i+2}+C'\frac{\Phi_0}{2\pi}{\ddot
\varphi'_{i+2}})
+(I'_c\sin\varphi'_{i+1}+C'\frac{\Phi_0}{2\pi}{\ddot
\varphi'_{i+1}})$. Using the equation of motion,
$C'_i(\Phi_0/2\pi)^2{\ddot\varphi}'_i=-\partial U_{\rm
eff}/\partial\varphi'_i$, obtained from the Lagrange equation, the
quantum Kirchhoff relation   reads
\begin{eqnarray}
\label{QK'}
-\frac{\Phi^2_0}{2\pi L_K}\frac{l}{2\pi}k'_i=\frac{\partial U_{\rm eff}}{\partial\varphi'_{i+2}}-\frac{\partial U_{\rm eff}}{\partial\varphi'_{i+1}}
-E'_{J}\sin\varphi'_{i+2}+E'_{J}\sin\varphi'_{i+1}.
\end{eqnarray}
We can confirm that the effective potential $U_{\rm eff}(\{\varphi_i,\varphi'_i\})$ in Eq. (\ref{Ufin}) also satisfies the quantum Kirchhoff relation
in Eq. (\ref{QK'}) with $k'_i$  in Eq. (\ref{k'i}).

\subsubsection*{Limiting case.}

In the system of Fig. \ref{scheme}(a) we can consider the limit that the length of branches goes to zero, $l'\rightarrow 0$,
and thus two nodes at the either ends of a branch collapse to a point. As a result, we have three loops with geometric inductance $L_s/3$
which meet at the trijunction. In this limit $L'_s \rightarrow 0$ and  $L'_{\rm eff}\approx L_s+9L'_s \rightarrow L_{\rm eff}\approx L_s$. Hence
the effective potential  $U_{\rm eff}(\{\varphi_i,\varphi'_i\})$ in Eq. (\ref{Ufin})  becomes
\begin{eqnarray}
U_{\rm eff}(\{\varphi_i,\varphi'_i\})&=&\frac{\Phi^2_0}{2(L_s/3)}\left[\left(m_1+f_1-\frac{\varphi_1+\varphi'_1}{2\pi}\right)^2+
\left( m_2+f_2-\frac{\varphi_2+\varphi'_2}{2\pi}\right)^2+\left(m_3+f_3-\frac{\varphi_3+\varphi'_3}{2\pi}\right)^2   \right]\nonumber\\
&-&\sum^3_{i=1}(E_{J}\cos\varphi_i+E'_{J}\cos\varphi'_i),
\end{eqnarray}
which describes the inductive energies of three loops with geometric inductance $L_s/3$  and the Josephson junction energies, \cite{Kim04,You,Qiu}
complying with the intuitive picture.

\subsection*{Circulator function.}

In order to perform the NISQ computing we need to construct a
scalable design with the circulator function, where the
trijunctions are connected to others and the current directions
can be controlled {\it in situ} in the circuit. However,  in the
design in Fig. \ref{scheme}(a) the trijunction is inside of the
loop so it is not possible to couple the branches with others
outside.  Hence we consider an improved design  where the
trijunction is located outside of the loop as shown in Fig.
\ref{scheme}(b).  In the Supplementary Information
we show an archetype for a scalable design.
Actually the inner branches and the trijunction
are turned over, but the design is topologically equivalent with
the design in Fig. \ref{scheme}(a). Here the length  $\tilde{l}$
of central branch is not equal with others anymore.

\begin{figure}[t]
\centering
\includegraphics[width=0.8\linewidth]{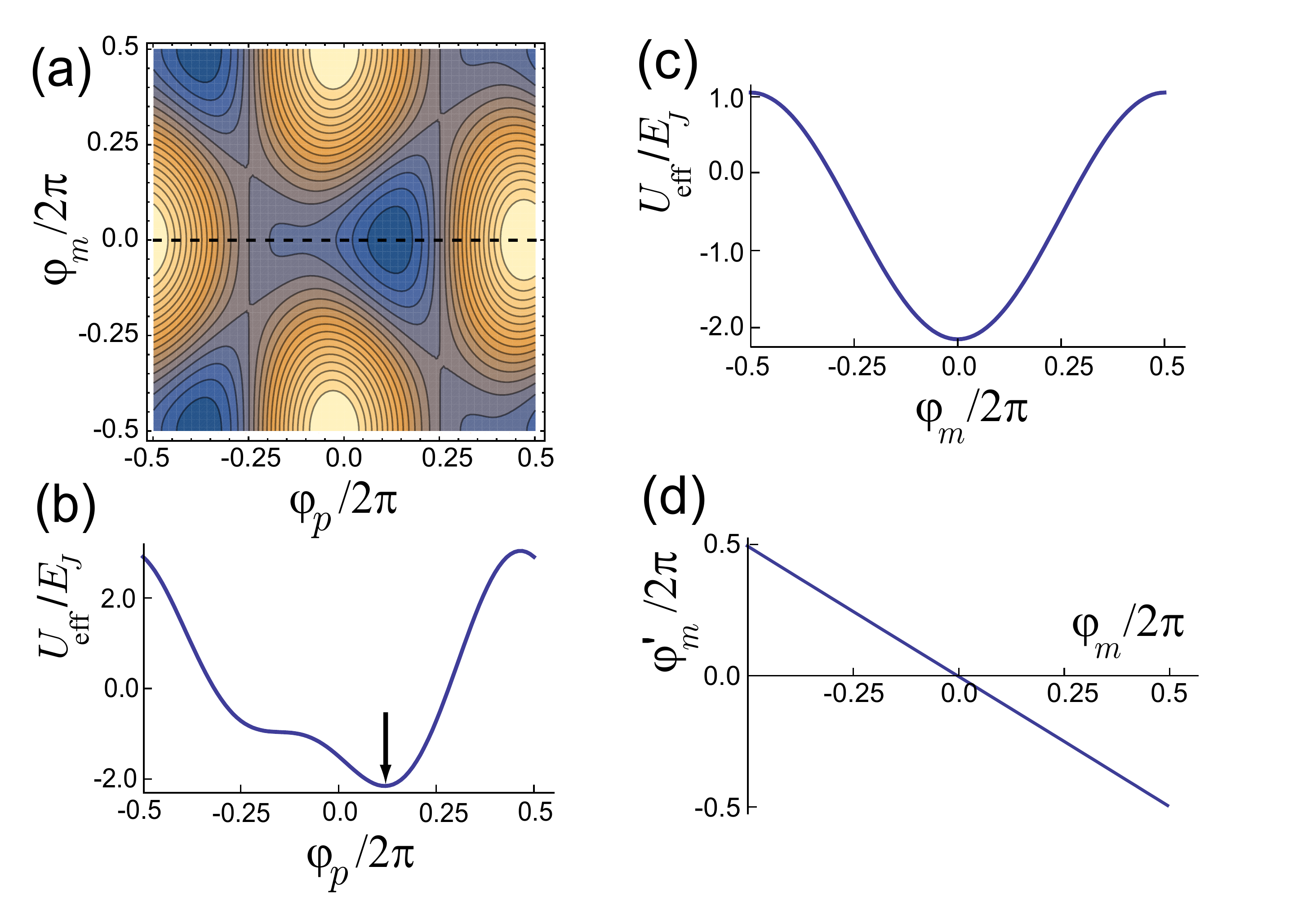}
\vspace{0.5cm}
\caption{(a) Contour plot for the effective potential $U_{\rm eff}$ for the system in Fig. \ref{scheme}(b)
as a function of $\varphi_p$ and $\varphi_m$ for $f_1=f=0.42,$ and $f_2=f_3=0$.
(b) Profile of $U_{\rm eff}$ along the dotted line in (a) for $\varphi_m=0$. At $\varphi_p/2\pi\approx 0.124$ $U_{\rm eff}$ has the minimum.
(c) The profile of $U_{\rm eff}$  for $\varphi_p/2\pi\approx 0.124$  shows $\varphi_m=0$ at the minimum of  $U_{\rm eff}$ .
(d) Plot of $\varphi'_m$ as a function of $\varphi_m$ which shows  $\varphi'_m=0$ at the minimum of
$U_{\rm eff}$ for $\varphi_p/2\pi\approx 0.124$ and $\varphi_m=0$  .}
\label{contour}
\end{figure}

We then introduce more general boundary conditions
for the scheme in Fig. \ref{scheme}(b) including the phase differences
across the Josephson junctions  as
\begin{eqnarray}
\label{center}
k'_2 l'-k'_3l'+k_1\frac{l}{3}+\varphi_1+\varphi'_1&=&2\pi(m_1+f_1-f_2-f_3+f_{\rm ind,1}),\\
\label{left}
-k'_3 l'+k'_1\tilde{l}-k_2\frac{l}{3}-\varphi_2-\varphi'_2&=&2\pi(-m_2-f_2+f_{\rm ind,2}),\\
\label{right}
k'_2 l'-k'_1\tilde{l}-k_3\frac{l}{3}-\varphi_3-\varphi'_3&=&2\pi(-m_3-f_3+f_{\rm ind,3}),
\end{eqnarray}
with integers $m_i$. The boundary condition in Eq. (\ref{center}) describes the
outmost loop containing the Josephson junctions with phase
differences $\varphi_1$ and $\varphi'_1$ and
the conditions in Eqs. (\ref{left}) and (\ref{right}) the left and right loop in Fig. \ref{scheme}(b).
With the geometric and kinetic inductances ${\tilde L}_s$ and ${\tilde L}_K=m_c{\tilde
l}/An_cq^2_c$ for the central branch, respectively, the induced fluxes  become
$f_{\rm ind,1}=-(1/2\pi)[(L'_s/L'_K)(k'_2-k'_3)l'+(L_s/L_K)k_1l/3],
f_{\rm ind,2}=-(1/2\pi)[-(L'_s/L'_K)k'_3l'+({\tilde L}_s/{\tilde L}_K)k'_1{\tilde l}-(L_s/L_K)k_2l/3]$ and
$f_{\rm ind,3}=-(1/2\pi)[(L'_s/L'_K)k'_2l'-({\tilde L}_s/{\tilde L}_K)k'_1{\tilde l}-(L_s/L_K)k_3l/3]$
to give rise to the relations similar to those in Eqs. (\ref{cbc}), (\ref{rbc}) and
(\ref{lbc}) where $k'_1l'$'s  are replaced with  $k'_1{\tilde l}$.
From these relations in conjunction with the relations in Eq.
(\ref{kcond}) we can similarly calculate $k_i$ and $k'_i$ with $i=1,2,3$ in
terms of $\varphi_i$ and $\varphi'_i$ (see the Supplementary Information).

In order to induce current flowing between the branches across $\varphi'_1$,
we initially apply the flux $\Phi_{\rm ext,1}$ so that $f_1=\Phi_{\rm ext,1}/\Phi_0=f$, but $f_2=f_3=0$.
We then can easily check that the following effective potential satisfies the equation of motion in Eqs. (\ref{QK}) and (\ref{QK'}),
\begin{eqnarray}
\label{Ueff-1}
U_{\rm eff}(\{\varphi_i,\varphi'_i\})&=&\frac{3\Phi^2_0}{4{\tilde L}_{\rm eff}}\left(-m_2+m_3+\frac{\varphi_2+\varphi'_2}{2\pi}-\frac{\varphi_3+\varphi'_3}{2\pi}\right)^2
+\frac{1}{2}\left(\frac{\Phi^2_0}{2L'_{\rm eff}}+\frac{\Phi^2_0}{L_{\rm eff}}\right)\left(n+f-\frac{\varphi_1+\varphi_2+\varphi_3}{2\pi}\right)^2\nonumber\\
&-&\frac{3\Phi^2_0}{2L'_{\rm eff}}\left(m_1+f-\frac{\varphi_1+\varphi'_1}{2\pi}\right)\left(n+f-\frac{\varphi_1+\varphi_2+\varphi_3}{2\pi}\right)+\frac{9\Phi^2_0}{4L'_{\rm eff}}\left(m_1+f-\frac{\varphi_1+\varphi'_1}{2\pi}\right)^2\\
&-&\sum^3_{i=1}(E_{J}\cos\varphi_i+E'_{J}\cos\varphi'_i),\nonumber
\end{eqnarray}
where ${\tilde L}_{\rm eff}\equiv L_K+L_s+3(L'_K+L'_s)+6({\tilde L}_K+{\tilde L}_s)$  is the effective  inductance of the central branch.
By manipulating the third term in Eq. (\ref{Ueff-1}) (see the Supplementary
Information) we can obtain the effective potential of the system in Fig. \ref{scheme}(b) as
\begin{eqnarray}
\label{Ueff-2}
U_{\rm eff}(\{\varphi_i,\varphi'_i\})\!\!\!&=&\!\!\!\frac{3\Phi^2_0}{2L'_{\rm eff}}\left(m_1\!+\!f\!-\!\frac{\varphi_1+\varphi'_1}{2\pi}\right)^2
\!\!\!+\!\!\frac32\left(\frac{\Phi^2_0}{2L'_{\rm eff}}\!+\!\frac{\Phi^2_0}{2{\tilde L}_{\rm eff}}\right)
\!\!\left[\left( m_2\!\!-\!\!\frac{\varphi_2+\varphi'_2}{2\pi}\right)^2\!\!+\!\!\left(m_3\!\!-\!\!\frac{\varphi_3+\varphi'_3}{2\pi}\right)^2   \right]\\
\!\!&+&\!\!\!\!\left(\frac{3\Phi^2_0}{2L'_{\rm eff}}\!-\!\frac{3\Phi^2_0}{2{\tilde L}_{\rm eff}}\right)\!\!\left( m_2\!-\!\frac{\varphi_2+\varphi'_2}{2\pi}\right)\!\!\left(m_3-\frac{\varphi_3+\varphi'_3}{2\pi}\right)\!\!+\!\!\left(\frac{\Phi^2_0}{2L_{\rm eff}}\!-\!\frac{\Phi^2_0}{2L'_{\rm eff}}\right)\left(n\!+\!f\!-\!\frac{\varphi_1+\varphi_2+\varphi_3}{2\pi}\right)^2\nonumber\\
&-&\sum_i(E_{Ji}\cos\varphi_i+E'_{Ji}\cos\varphi'_i).\nonumber
\end{eqnarray}
If we consider that the inductances of left, right and central
branches are all equal, ${\tilde l}=l'$, ${\tilde
L}_s=L'_s$, ${\tilde L}_K=L'_K$, and thus ${\tilde L}_{\rm
eff}=L'_{\rm eff}$, the effective potential $U_{\rm
eff}(\{\varphi_i,\varphi'_i\})$ in Eq. (\ref{Ueff-2}) can be
reduced to $U_{\rm eff}(\{\varphi_i,\varphi'_i\})$ in Eq.
(\ref{Ufin}) for the system in Fig. \ref{scheme}(a) with $f_1=f$
and $f_2=f_3=0$.  Figure  \ref{contour} shows the effective potential for the design in Fig. \ref{scheme}(b),
which is qualitatively similar to that for the model in Fig. \ref{scheme}(a).

We introduce a coordinate transformation such as
$\varphi_p=(\varphi_2+\varphi_3)/2, \varphi_m=(\varphi_2-\varphi_3)/2, \varphi'_p=(\varphi'_2+\varphi'_3)/2,$ and $ \varphi'_m=(\varphi'_2-\varphi'_3)/2$. The effective potential in Eq. (\ref{Ueff-2}), then, can be expressed as
\begin{eqnarray}
U_{\rm eff}(\varphi_p,\varphi_m,\varphi'_p,\varphi'_m,\varphi_1)\!\!\!\!\!\!&=&\!\!\!\!\!\!
\frac{3\Phi^2_0}{2L'_{\rm eff}}\!\left(\!m_1\!-\!n'\!+\!f\!-\!\frac{\varphi_1\!-\!2\varphi'_p}{2\pi}\right)^2
+\left(\frac{\Phi^2_0}{2L_{\rm eff}}-\frac{\Phi^2_0}{2L'_{\rm eff}}\right)\left(n+f-\frac{\varphi_1+2\varphi_p}{2\pi}\right)^2\nonumber\\
\!\!\!\!\!\!&+&\!\!\!\!\!\frac32\left(\frac{\Phi^2_0}{2L'_{\rm eff}}\!+\!\frac{\Phi^2_0}{2{\tilde L}_{\rm eff}}\right)
\left[\left(\! m_2\!-\!\frac{\varphi_p\!+\!\varphi_m\!+\!\varphi'_p\!+\!\varphi'_m}{2\pi}\right)^2\!+\!\left(\!m_3\!-\!\frac{\varphi_p\!-\!\varphi_m\!+\!\varphi'_p\!-\!\varphi'_m}{2\pi}\right)^2 \right]\nonumber\\
\!\!\!\!\!\!&+&\!\!\!\!\!\!\frac32\left(\frac{\Phi^2_0}{2L'_{\rm eff}}\!-\!\frac{\Phi^2_0}{2{\tilde L}_{\rm eff}}\right)
\left(\! m_2\!-\!\frac{\varphi_p\!+\!\varphi_m\!+\!\varphi'_p\!+\!\varphi'_m}{2\pi}\right)\left(\!m_3\!-\!\frac{\varphi_p\!-\!\varphi_m\!+\!\varphi'_p\!-\!\varphi'_m}{2\pi}\right)\\
\!\!\!\!\!\!&-&\!\!\!\!\!\!E_{J}\cos\varphi_1-2E_{J}\cos\varphi_p\cos\varphi_m-E'_{J}\cos2\varphi'_p-2E'_{J}\cos\varphi'_p\cos\varphi'_m, \nonumber
\end{eqnarray}
where we use $\varphi'_1=2\pi n'-(\varphi'_2+\varphi'_3)=2\pi
n'-2\varphi'_p$. Figure \ref{contour}(a) shows the effective
potential $U_{\rm eff}$ as a function of $(\varphi_p,\varphi_m)$
for $m_1=m_2=m_3=n=n'=0$, which is minimized with respect to
$\varphi'_p, \varphi'_m$ and $\varphi_1$. If the value of the
external flux $f=0.5$,  two degenerate current
states, clockwise and counterclockwise, are  superposed
so that we cannot determine the direction of current. We thus set
the value of the external flux $f=0.42$ to obtain a stable
minimum. The effective potential  $U_{\rm eff}(\varphi_p,\varphi_m)$ along the dotted line
in Fig. \ref{contour}(a) is shown in Fig. \ref{contour}(b), where
$U_{\rm eff}(\varphi_p,\varphi_m)$ has a  minimum at
 $\varphi_p/2\pi\approx 0.124$. Figure \ref{contour}(c)
shows the profile of  effective potential $U_{\rm eff}(\varphi_p,\varphi_m)$ as a
function of $\varphi_m$ for $\varphi_p/2\pi\approx 0.124$. Here
the effective potential has the minimum at  $\varphi_m=0, i.
e.,  \varphi_2=\varphi_3$. Figure \ref{contour}(d) show that
$\varphi'_m=0, i. e.,  \varphi'_2=\varphi'_3$ at the minimum of
the effective potential $U_{\rm eff}(\varphi_p,\varphi_m)$.
From Eqs. (\ref{Ik}) and (\ref{ki}) we can see that $k_2=k_3$ and
thus $I_2=I_3$ and from Eq. (\ref{k'i}) $k'_1=0$, and thus
$I'_1=0$, which is consistent with the current conservations,
$I_3-I_2=I'_1=0$, in Eq. (\ref{kcond}). Hence, in Fig.
\ref{scheme}(b)  we can determine the direction of current such as
$I'_3=-I'_2\neq 0$ , and $I'_1=0$. If we consider the case that
$f_3=f, f_1=f_2=0$ or $f_2=f, f_1=f_3=0$, the currents become
$I'_2=-I'_1\neq 0$, $I'_3=0$ or
 $I'_3=-I'_1\neq 0$, $I'_2=0$, respectively.
Hence we can selectively determine the direction of currents flowing through a trijunction
by threading a magnetic flux into a specific loop in the design of Fig. \ref{scheme}(b),
which can realize the circulator function in a scalable design.



\subsection*{Braiding of Majorana zero modes.}

\begin{figure}[t]
\centering
\vspace{-0cm}
\hspace{-0cm}
\includegraphics[width=0.7\linewidth]{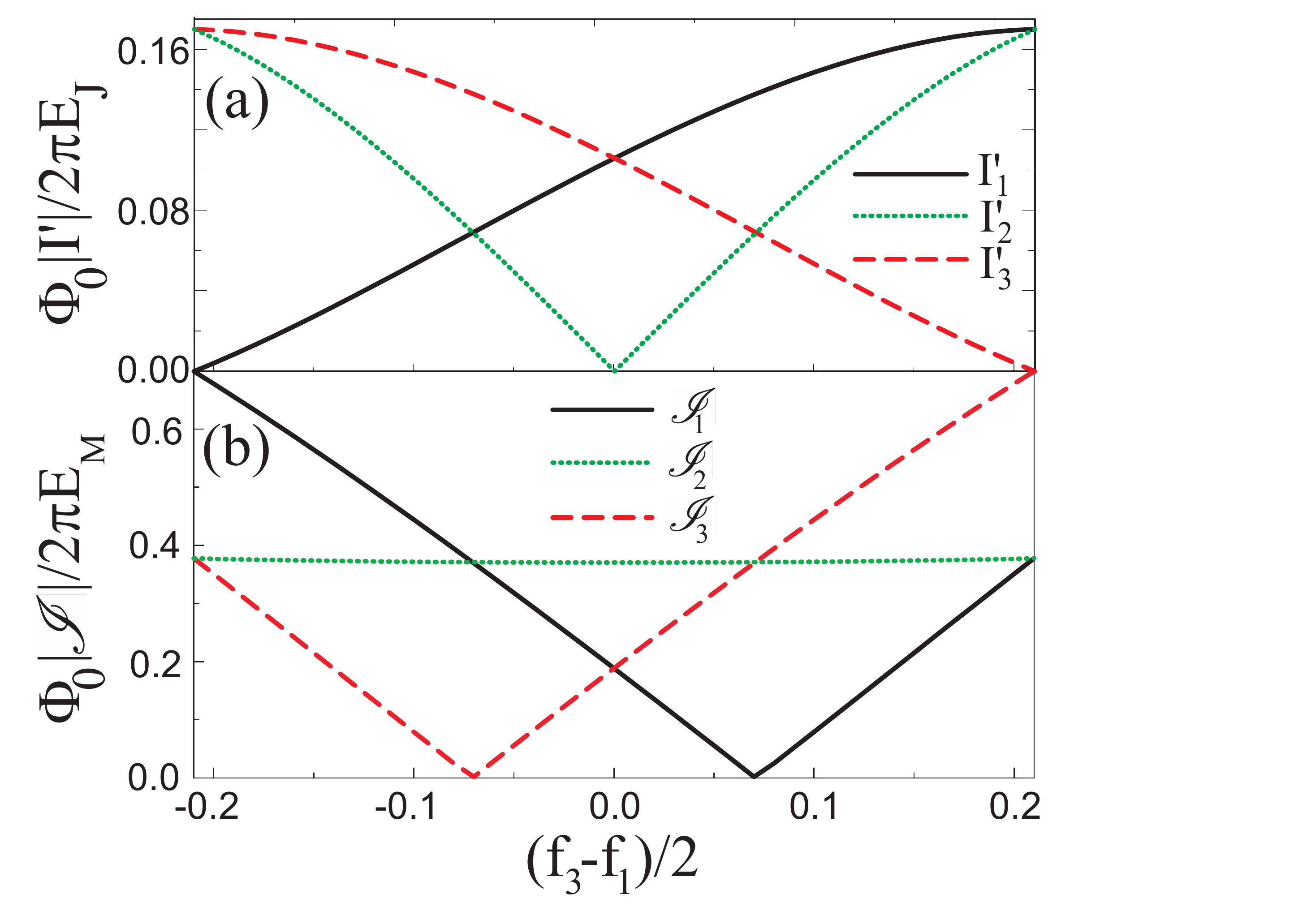}
\vspace{0.5cm} \caption{(a) Currents $I'_i$ in branches as a
function of $(f_3-f_1)/2$. When $f_1$ starts from $f_1=0.42$ with
$f_2=f_3=0$, the currents $|I'_1|=|I'_3|\neq 0$ with $|I'_1|$=0.
As $f_3$  increases while $f_1$ decreases to zero, the current
flow changes so that   $|I'_1|=|I'_2|\neq 0$ with $|I'_3|$=0. (b)
Currents ${\cal I}_i$ carried through MZMs across trijunction.
For $f_1=0.42$ and $f_2=f_3=0$ the current ${\cal I}_1$ has larger
amplitude than $|{\cal I}_2|=|{\cal I}_3|$, but for $f_3=0.42$ and
$f_1=f_2=0$, $|{\cal I}_3|$ becomes larger, so the asymmetry is
changed.} 
\label{IPFig}
\end{figure}

\begin{figure}[t]
\centering
\vspace{-0cm}
\hspace{-0cm}
\includegraphics[width=0.7\linewidth]{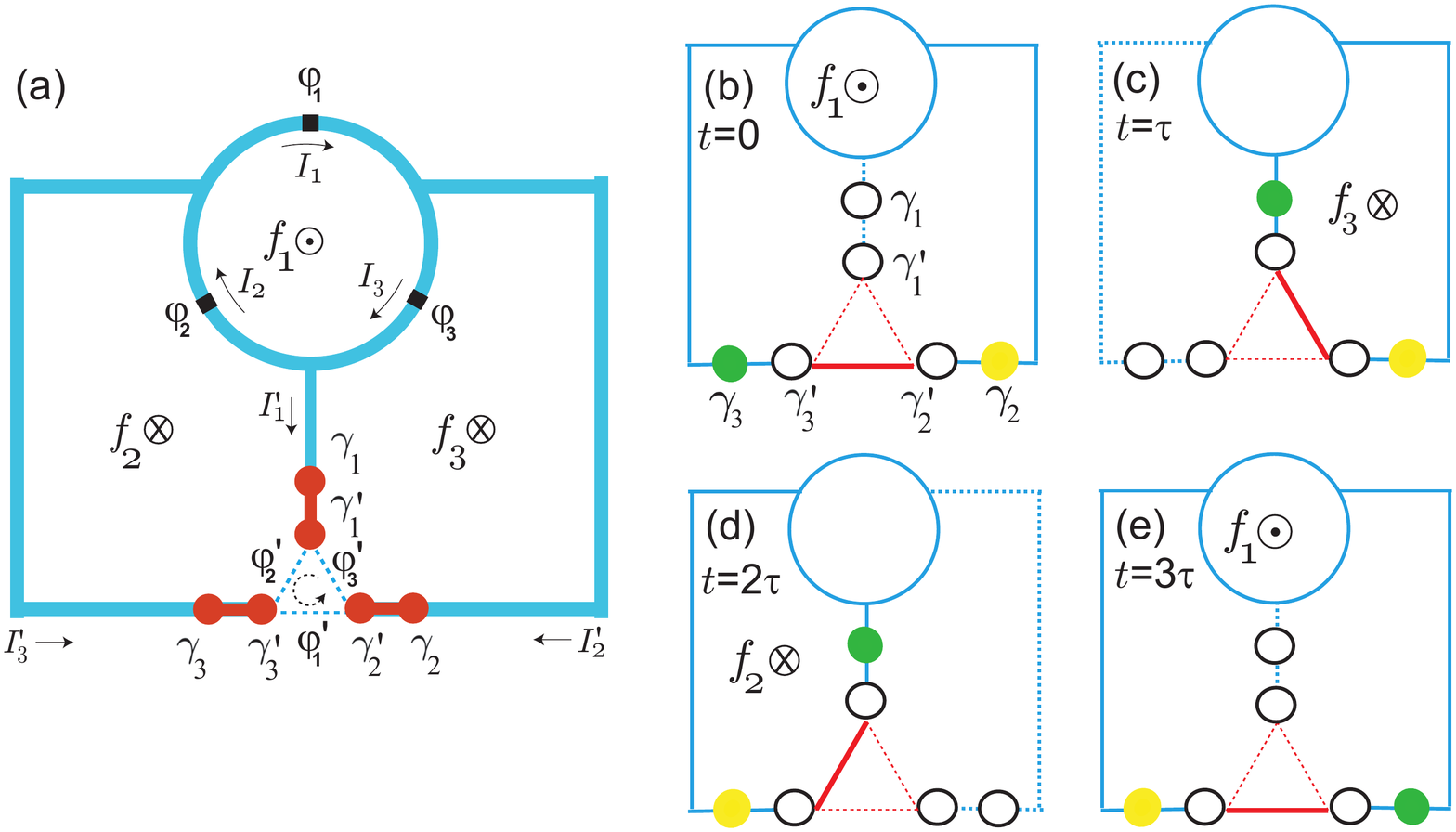}
\vspace{-1.0cm}
\caption{(a) Three MZM (red circle) pairs are
introduced at the end of branches where three MZMs,
$\gamma'_i$, are coupled through  a Josephson trijunction.
Braiding sequence of system in (a): by applying adiabatically the
fluxes (b)$f_1,$ (c)$f_3$, (d)$f_2$, and finally (e)$f_1$ again,
the green and yellow MZMs are exchanged with each other to
complete a non-Abelian braiding procedure. In the  branches
represented as dotted line there is no current flowing. In trijunction
thick red line corresponds to a large current amplitude of ${\cal I}_i$.}  
\label{braid}
\end{figure}

\begin{figure}[b]
\vspace{-1.0cm}
\centering
\includegraphics[width=0.9\linewidth]{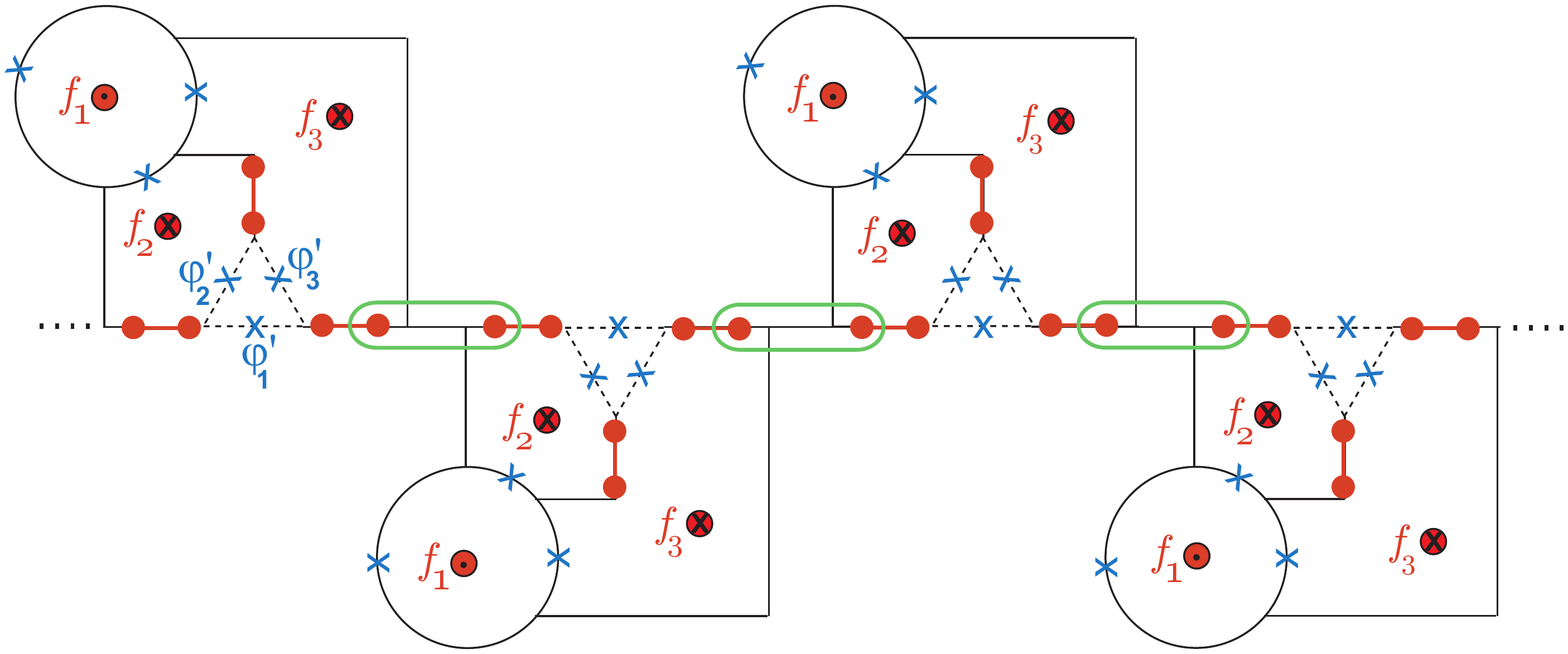}
\vspace{-5cm}
\caption{A scalable design for a superconducting circuit with MZMs. Two MZMs in each circuit of Fig. \ref{braid}(a) can be coupled in the green box to form a one-dimensional lattice structure.}
\label{scale}
\end{figure}

We can use the circulator function for the braiding of Majorana zero modes (MZM) for topological quantum computing.
As shown in Fig. \ref{braid}(a) we  introduce three pairs of MZMs in the semiconducting nanowire
with  p-wave-like superconductivity  induced  from s-wave superconducting branch via proximity effect.
For the quantum computing  the scheme for quantum gate operation should be provided. Hence we consider
the system of Fig. \ref{scheme}(b) because for the system of Fig. \ref{scheme}(a) the MZMs are inside of the loop so that the MZMs cannot interact with MZMs outside.\cite{Stenger}


In Fig. \ref{IPFig}(a) we show the currents  $I'_1=I_3-I_2$, $I'_2=I_1-I_3,$ and $I'_3=I_2-I_1$
of the system in Fig. \ref{scheme}(b) as a  function of $f_1-f_2$.
If $f_1=f=0.42$ with $f_2=f_3=0$,   the current direction is determined
such that  $I'_1=0$, but $I'_2=I'_3\neq 0$.
In this case the current flows between the branch
with  $\gamma_2$ and the branch with $\gamma_3$.
This is the initial state of the system shown in Fig. \ref{braid}(b), where
the three MZMs, $\gamma'_1,\gamma'_2$ and $\gamma'_3$, are tunnel-coupled with each other
through the Hamiltonian  \cite{Heck,Stenger}
\begin{eqnarray}
\label{HT}
H_T=iE_M(\gamma'_1\gamma'_2\cos\frac{\varphi'_3}{2}+\gamma'_2\gamma'_3\cos\frac{\varphi'_1}{2}
+\gamma'_3\gamma'_1\cos\frac{\varphi'_2}{2})+i\alpha\sum^3_{i=1}\gamma_i\gamma'_i
\end{eqnarray}
with Majorana Josephson energy $E_M$ and coupling energy $\alpha$.
Then the current carried through MZMs across trijunction is given by
\begin{eqnarray}
\label{Imzm}
{\cal I}_{i}=\frac{2e}{\hslash}\frac{\partial}{\partial \varphi'_i}H_T=-\frac{2\pi E_M}{\Phi_0}i\gamma'_{i+1}\gamma'_{i+2}\sin\frac{\varphi'_i}{2}
\end{eqnarray}
with a  $4 \pi$-periodic behavior. \cite{Kayyalha}
Actually we have $\varphi'_1/2\pi\approx 0.246$ and
$\varphi'_2/2\pi=\varphi'_3/2\pi\approx -0.123$  at the minimum of the effective potential
$U_{\rm eff}(\varphi_p,\varphi_m)$ in Fig. \ref{contour}(a).
Then the current   ${\cal I}_1$ has a larger amplitude than  ${\cal I}_2={\cal I}_3$
as shown in Fig. \ref{IPFig}(b), which is denoted as a solid (dotted) line for  ${\cal I}_1 ({\cal I}_2$ and ${\cal I}_3)$
in the trijunction of  Fig. \ref{braid}(b).
As shown in Eq. (\ref{Imzm}) the current mediated by MZMs ${\cal
I}_i\propto \sin\varphi'_i/2$, while the Cooper pair current
${\tilde I}_i\propto \sin\varphi'_i$. If we consider a simplified
model such that the Josephson junctions in the three-junction loop
in Fig. \ref{scheme}(a) are removed as in the previous study,
\cite{Stenger} the boundary condition becomes approximately
$\varphi'_i-2\pi f_i\approx 0$. Here, even if we set $f_i= 0.5$
and thus $\varphi'_i\approx \pi$, we cannot switch off the current
mediated by MZMs as ${\cal I}_i \neq 0$ while  ${\tilde I}_i\approx 0$.
Hence, instead of switching off  ${\cal I}_i$ we change the
current direction by using circulating function to perform the
braiding operation.

In general, for $f_i=0.42$ with $f_{i\pm1}=0$  we have  $\varphi'_i/2\pi\approx 0.246$ and
$\varphi'_{i\pm1}/2\pi\approx -0.123$.
The different phases are due to the current direction, resulting
in the asymmetry in the amplitude of ${\cal I}_i$ at the trijunction.
In next stage we adiabatically apply the flux $f_3$, while decreasing $f_1$ (See Eq. (S27) of Supplementary  Information for general $f_i$).
In Fig. \ref{IPFig}(a), then, $|I'_1|$ increases while $|I'_3|$ decreases. In the meanwhile, $|I'_2|$ decreases to zero and then
grows up to the maximum value.   Finally for $f_3=0.42$ with $f_1=f_2=0$, we have
$I'_3=0$, but $I'_1=I'_2\neq 0$. Hence the current direction is changed:
the current ${\cal I}_i$ flows between the branch with $\gamma_1$ and the branch with $\gamma_2$
but there is no current in the branch with $\gamma_3$ as shown in Fig. \ref{braid}(c),
and meanwhile the green MZM loses its weight in $\gamma_3$ and
gains weight in $\gamma_1$.  Here the current  ${\cal I}_3$ has a larger amplitude than  ${\cal I}_1={\cal I}_2$,
and thus the asymmetry in the amplitude of ${\cal I}_i$ is changed.
In this way, between $t=\tau$ and $t=2\tau$,   the yellow MZM loses its
weight in $\gamma_2$ and gains weight in $\gamma_3$ as shown in Fig. \ref{braid}(d).
At the last stage  the green MZM loses its weight in $\gamma_1$ and gains weight in $\gamma_2$.
As a result, the green and yellow MZMs are exchanged with each other as shown in Fig. \ref{braid}(e),
completing the braiding operation.
%

In Fig. \ref{scale} we show an architecture for a scalable design
for a superconducting circuit with MZMs. Two MZMs belong to
different trijunctions  (the green box in Fig. \ref{scale})  can
be coupled or fused to perform  quantum gate operations and
quantum measurements. For the green box operation, for example, we
can introduce a  gate voltage applied to the sector between two
MZMs to control the chemical potential of the nanowire.
\cite{Alicea} Though the system in  Fig. \ref{scale} is
one-dimensional, we can extend it to two-dimensional lattice
straightforwardly.

\section*{Discussion}

In conclusion, we proposed a scheme for the  circulator function
in a superconducting circuit consisting of three small loops and
branches which meet at a trijunction. Usually the effective
potential in the Hamiltonian for superconducting circuit is
phenomenologically obtained. However in this study  we obtained
the boundary conditions from the fundamental  fluxoid quantization
condition for the superconducting loop to  derive the effective
potential of the system analytically, which is required for
accurate and systematic study for the quantum information
processing applications. We expect that this kind  of study can be
applied to other systems.

At the minimum of the effective potential we can see that two branches carry current while the
other does not. By applying a magnetic flux into one of the loops
we can determine which branches among three carry the current,
achieving the circulator function. For the NISQ computing we need
to perform the circulator function in a scalable design. We thus
introduced an improved model where the trijunction is extracted
out from the outmost loop to interact with other external
branches.   For the improved design we obtained the ground state
of the system from the effective potential, and showed that it can
perform the circulator function in the trijunction loop.

Instead of switching off  the current mediated by MZMs in the previous study, in this
study we selectively choose the current directions  to give rise
to MZM braiding. We thus use the circulator function to achieve  a
non-Abelian braiding operation by introducing three pairs of MZMs
in the branches that meet at a trijunction  in the improved model
where MZMs are introduced outside of the loop.  The circulator
function determines the phases of the trijunction and thus the
coupling between the MZMs. Initially we apply a magnetic flux into
one of the three loops to selectively couple two pairs of MZMs. By
applying adiabatically a flux into another loop while decreasing
the previous flux we are able to gain the weight of  MZM while
losing in the previous branch. Consecutive executions in this way
can perform the braiding operation between two MZMs. This scheme
could be extended to a scalable design to implement braiding
operations in one- or two-dimensional circuits.


\section*{Acknowledgements}

This research was  supported by Basic Science Research Program through the National Research Foundation of Korea(NRF)
funded by the Ministry of Education(2019R1I1A1A01061274), 2020 Hongik University Research Fund,
and  Korea Institute for Advanced Study(KIAS) grant funded by the Korea government.

\section*{Author contributions statement}

M.D.K. solely developed the ideas, performed  calculations,
and wrote the manuscript.

\section*{Competing interests}

The author declares no competing interests.

\section*{Additional information}


\textbf{Correspondence} and requests for materials should be addressed to M.D.K.







\newpage

{\Large Supplementary Information for "Circulator function  in a Josephson junction circuit and  braiding of Majorana zero modes"}

\section{Effective potential of the improved circuit }

\begin{figure}[t]
\centering
\hspace{-0cm}
\includegraphics[width=1.0\linewidth]{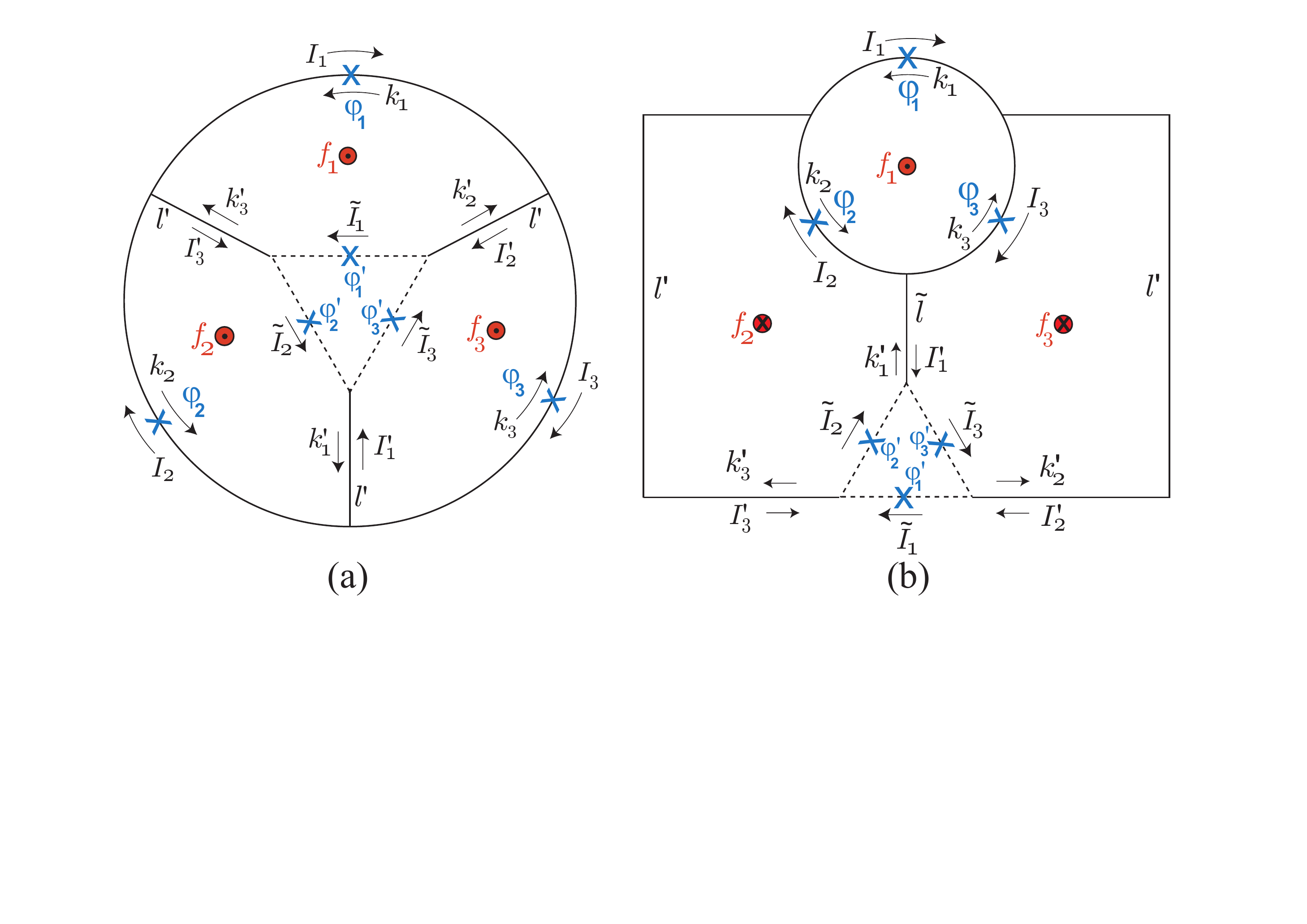}
\vspace{-4cm}
\caption{Simplified picture of Fig. 1 in the main manuscript.}
\label{figS}
\end{figure}

The design of Fig. 1 in the main manuscript can be simplified as
 Fig. \ref{figS}. In the figures we denote the currents $I_i$ and $I'_i$ in the loop
whose direction is  opposite to the Cooper pair  wave vector $k_i$ and $k'_i$, respectively.
In this Supplementary Information we consider the more general case of
Fig. \ref{figS}(b). Here, we consider that $f_1=f$ and $f_2=f_3=0$.
The boundary conditions  for the scheme in Fig. \ref{figS}(b) including the phase differences
across the Josephson junctions are represented as,
\begin{eqnarray}
\label{center}
k'_2 l'-k'_3l'+k_1\frac{l}{3}+\varphi_1+\varphi'_1&=&2\pi(m_1+f+f_{\rm ind,1}),\\
\label{left}
-k'_3 l'+k'_1\tilde{l}-k_2\frac{l}{3}-\varphi_2-\varphi'_2&=&2\pi(-m_2+f_{\rm ind,2}),\\
\label{right}
k'_2 l'-k'_1\tilde{l}-k_3\frac{l}{3}-\varphi_3-\varphi'_3&=&2\pi(-m_3+f_{\rm ind,3}),
\end{eqnarray}
with integers $m_i$.

Equation (\ref{center}) describes the boundary condition for the
outmost loop containing the Josephson junctions with phase
differences $\varphi_1$ and $\varphi'_1$, and Eqs. (\ref{left})
and (\ref{right})  the left and right loop in Fig. \ref{figS}(b).
The induced flux, $f_{\rm ind,i}=\Phi_{\rm ind,i}/\Phi_0$, can be
written as
\begin{eqnarray}
f_{\rm ind,1}&=&(1/\Phi_0)(I_1+I_2+I_3)L_s/3,\\
f_{\rm ind,2}&=&(1/\Phi_0)(-L'_sI'_3+\tilde{L}_sI'_1-L_sI_2/3),\\
f_{\rm ind,3}&=&(1/\Phi_0)(L'_sI'_2-\tilde{L}_sI'_1-L_sI_3/3),
\end{eqnarray}
where the  Cooper pair current $I$ is given by
\begin{eqnarray}
\label{Ik}
I_i=-(n_cAq_c/m_c)\hslash k_i.
\end{eqnarray}
With the kinetic inductances of side branches, central branch, and the three-Josephson junction loop being
$L'_K=m_cl'/An_cq^2_c, {\tilde L}_K=m_c{\tilde l}/An_cq^2_c$ and $L_K=m_cl/An_cq^2_c$, respectively, the induced fluxes  become
\begin{eqnarray}
f_{\rm ind,1}&=&-\frac{1}{2\pi}[(L'_s/L'_K)(k'_2-k'_3)l'+(L_s/L_K)k_1l/3],\\
f_{\rm ind,2}&=&-\frac{1}{2\pi}[-(L'_s/L'_K)k'_3l'+({\tilde L}_s/{\tilde L}_K)k'_1{\tilde l}-(L_s/L_K)k_2l/3],\\
f_{\rm ind,3}&=&-\frac{1}{2\pi}[(L'_s/L'_K)k'_2l'-({\tilde L}_s/{\tilde L}_K)k'_1{\tilde l}-(L_s/L_K)k_3l/3].
\end{eqnarray}
Then the boundary conditions are represented as
\begin{eqnarray}
\label{cbc}
\left(1+\frac{L_s}{L_K}\right)k_1\frac{l}{3}+\left(1+\frac{L'_s}{L'_K}\right)k'_2l'-\left(1+\frac{L'_s}{L'_K}\right)k'_3l'
&=&2\pi\left(m_1+f-\frac{\varphi_1+\varphi'_1}{2\pi}\right)\\
\label{rbc}
\left(1+\frac{L_s}{L_K}\right)k_2\frac{l}{3}+\left(1+\frac{L'_s}{L'_K}\right)k'_3l'-\left(1+\frac{{\tilde L}_s}{{\tilde L}_K}\right)k'_1{\tilde l}
&=&2\pi\left(m_2-\frac{\varphi_2+\varphi'_2}{2\pi}\right)\\
\label{lbc}
\left(1+\frac{L_s}{L_K}\right)k_3\frac{l}{3}+\left(1+\frac{{\tilde L}_s}{{\tilde L}_K}\right)k'_1{\tilde l}-\left(1+\frac{L'_s}{L'_K}\right)k'_2l'
&=&2\pi\left(m_3-\frac{\varphi_3+\varphi'_3}{2\pi}\right)
\end{eqnarray}
with ${\tilde L}_s$ and $L'_s$ being the self  inductance of the central  and the side branch, respectively.

The current conservation conditions, $I_1=I_3+I'_2, I_2=I_1+I'_3$, and
$I_3=I_2+I'_1$,  at the nodes of three-Josephson junction loop
give rise to the relations,
\begin{eqnarray}
\label{ccc}
k_1=k_3+k'_2, k_2=k_1+k'_3, k_3=k_2+k'_1.
\end{eqnarray}


From Eqs. (\ref{cbc}), (\ref{rbc}), (\ref{lbc}), and  (\ref{ccc}) we can obtain
\begin{eqnarray}
\label{k1}
k_1&=&\frac{2\pi}{l}\frac{3L_K}{L'_{\rm eff}}\left(m_1+f-\frac{\varphi_1+\varphi'_1}{2\pi}\right)
+\frac{2\pi}{l}\left(\frac{L_K}{L_{\rm eff}}-\frac{L_K}{L'_{\rm eff}}\right)\left(n+f-\frac{\varphi_1+\varphi_2+\varphi_3}{2\pi}\right),\\
\label{k2}
k_2&=&-\frac{\pi}{l}\frac{3L_K}{L'_{\rm eff}}\left(m_1+f-\frac{\varphi_1+\varphi'_1}{2\pi}\right)
-\frac{\pi}{l}\frac{3L_K}{\tilde{L}_{\rm eff}}\left(m+\frac{\varphi_2+\varphi'_2}{2\pi}-\frac{\varphi_3+\varphi'_3}{2\pi}\right)\nonumber\\
&+&\frac{2\pi}{l}\left(\frac{L_K}{L_{\rm eff}}+\frac{L_K}{2L'_{\rm eff}}\right)\left(n+f-\frac{\varphi_1+\varphi_2+\varphi_3}{2\pi}\right),\\
\label{k3}
k_3&=&-\frac{\pi}{l}\frac{3L_K}{L'_{\rm eff}}\left(m_1+f-\frac{\varphi_1+\varphi'_1}{2\pi}\right)
+\frac{\pi}{l}\frac{3L_K}{\tilde{L}_{\rm eff}}\left(m+\frac{\varphi_2+\varphi'_2}{2\pi}-\frac{\varphi_3+\varphi'_3}{2\pi}\right)\nonumber\\
&+&\frac{2\pi}{l}\left(\frac{L_K}{L_{\rm eff}}+\frac{L_K}{2L'_{\rm eff}}\right)\left(n+f-\frac{\varphi_1+\varphi_2+\varphi_3}{2\pi}\right),\\
k'_1&=&\frac{2\pi}{l}\frac{3L_K}{\tilde{L}_{\rm eff}}\left(m+\frac{\varphi_2+\varphi'_2}{2\pi}-\frac{\varphi_3+\varphi'_3}{2\pi}\right),\\
k'_2&=&-\frac{\pi}{l}\frac{3L_K}{\tilde{L}_{\rm eff}}\left(m+\frac{\varphi_2+\varphi'_2}{2\pi}
-\frac{\varphi_3+\varphi'_3}{2\pi}\right)\nonumber\\
&-&\frac{\pi}{l}\frac{9L_K}{L'_{\rm eff}}\left[m_1-f+\frac{\varphi_1+\varphi'_1}{2\pi}+\frac13\left(n+f-\frac{\varphi_1+\varphi_2+\varphi_3}{2\pi}\right)\right],\\
k'_3&=&-\frac{\pi}{l}\frac{3L_K}{\tilde{L}_{\rm eff}}\left(m+\frac{\varphi_2+\varphi'_2}{2\pi}
-\frac{\varphi_3+\varphi'_3}{2\pi}\right)\nonumber\\
&+&\frac{\pi}{l}\frac{9L_K}{L'_{\rm eff}}\left[m_1-f+\frac{\varphi_1+\varphi'_1}{2\pi}+\frac13\left(n+f-\frac{\varphi_1+\varphi_2+\varphi_3}{2\pi}\right)\right],
\end{eqnarray}
where $L_{\rm eff}= L_K+L_s, L'_{\rm eff}= L_K+L_s+9(L'_K+L'_s)$,
and ${\tilde L}_{\rm eff}\equiv L_K+L_s+3(L'_K+L'_s)+6({\tilde
L}_K+{\tilde L}_s)$ are the effective inductances of
three-Josephson junction loop, side branches, and central branch,
respectively.


By using the quantum Kirchhoff relation  the equation of motion  can be represented as
\begin{eqnarray}
\label{QK1}
&&\frac{\Phi^2_0}{2\pi L_K}\frac{l}{2\pi}k_i-E_{J}\sin\phi_i=-\frac{\partial U_{\rm eff}}{\partial\phi_i},\\
\label{QK2}
&&-\frac{\Phi^2_0}{2\pi L_K}\frac{l}{2\pi}k'_i=\frac{\partial U_{\rm eff}}{\partial\varphi'_{i+2}}-\frac{\partial U_{\rm eff}}{\partial\varphi'_{i+1}}
-E'_{J}\sin\varphi'_{i+2}+E'_{J}\sin\varphi'_{i+1}.
\end{eqnarray}
We then can obtain the effective potential satisfying Eqs. (\ref{QK1}) and (\ref{QK2}) as  follows,
\begin{eqnarray}
\label{Ueff}
U_{\rm eff}(\{\varphi_i,\varphi'_i\})&\!\!=\!\!&\frac{3\Phi^2_0}{4{\tilde L}_{\rm eff}}\left(\!\!-m_2\!+\!m_3\!+\!\frac{\varphi_2\!+\!\varphi'_2}{2\pi}\!-\!\frac{\varphi_3\!+\!\varphi'_3}{2\pi}\right)^2
\!\!+\!\frac{1}{2}\left(\frac{\Phi^2_0}{2L'_{\rm eff}}\!+\!\frac{\Phi^2_0}{L_{\rm eff}}\right)\left(\!n\!+\!f\!-\!\frac{\varphi_1\!+\!\varphi_2\!+\!\varphi_3}{2\pi}\right)^2\nonumber\\
&\!-\!&\frac{3\Phi^2_0}{2L'_{\rm eff}}\left(\!m_1\!+\!f\!-\!\frac{\varphi_1\!+\!\varphi'_1}{2\pi}\right)\!\left(\!n\!+\!f\!-\!\frac{\varphi_1\!+\!\varphi_2\!+\!\varphi_3}{2\pi}\right)\!+\!\frac{9\Phi^2_0}{4L'_{\rm eff}}\left(\!m_1\!+\!f\!-\!\frac{\varphi_1\!+\!\varphi'_1}{2\pi}\right)^2\nonumber\\
&-&\sum_i(E_{Ji}\cos\varphi_i+E'_{Ji}\cos\varphi'_i).
\end{eqnarray}
The third term of Eq. (\ref{Ueff}) can be rewritten as
\begin{eqnarray}
\label{3rd}
&&\frac{3\Phi^2_0}{4L'_{\rm eff}}\left[\left(n+f-\frac{\varphi_1+\varphi_2+\varphi_3}{2\pi}\right)-\left(m_1+f-\frac{\varphi_1+\varphi'_1}{2\pi}\right)\right]^2\nonumber\\
&&-\frac{3\Phi^2_0}{4L'_{\rm eff}}\left(n+f-\frac{\varphi_1+\varphi_2+\varphi_3}{2\pi}\right)^2-\frac{3\Phi^2_0}{4
L'_{\rm eff}}\left(m_1+f-\frac{\varphi_1+\varphi'_1}{2\pi}\right)^2,
\end{eqnarray}
where  by using $\varphi'_1=2\pi m'-\varphi'_2-\varphi'_3$  and choosing appropriate $m'$
the first term of Eq. (\ref{3rd})  can be represented as
\begin{eqnarray}
\frac{3\Phi^2_0}{4L'_{\rm eff}}\left(m_2-\frac{\varphi_2+\varphi'_2}{2\pi}
+m_3-\frac{\varphi_3+\varphi'_3}{2\pi}\right)^2.
\end{eqnarray}
As a result, the effective potential $U_{\rm eff}(\{\varphi_i,\varphi'_i\})$ in Eq. (\ref{Ueff}) is reexpressed as follows,
\begin{eqnarray}
U_{\rm eff}(\{\varphi_i,\varphi'_i\})\!\!\!&=&\!\!\!\frac{3\Phi^2_0}{2L'_{\rm eff}}\left(m_1\!+\!f\!-\!\frac{\varphi_1+\varphi'_1}{2\pi}\right)^2
\!\!\!+\!\!\frac32\left(\frac{\Phi^2_0}{2L'_{\rm eff}}\!+\!\frac{\Phi^2_0}{2{\tilde L}_{\rm eff}}\right)
\!\!\left[\left( m_2\!\!-\!\!\frac{\varphi_2+\varphi'_2}{2\pi}\right)^2\!\!+\!\!\left(m_3\!\!-\!\!\frac{\varphi_3+\varphi'_3}{2\pi}\right)^2   \right]\nonumber\\
&+&\left(\frac{\Phi^2_0}{2L_{\rm eff}}\!-\!\frac{\Phi^2_0}{2L'_{\rm eff}}\right)\left(n\!+\!f\!-\!\frac{\varphi_1+\varphi_2+\varphi_3}{2\pi}\right)^2-\sum_i(E_{Ji}\cos\varphi_i+E'_{Ji}\cos\varphi'_i).\nonumber\\
\!&+&\left(\frac{3\Phi^2_0}{2L'_{\rm eff}}\!-\!\frac{3\Phi^2_0}{2{\tilde L}_{\rm eff}}\right)\!\!\left( m_2\!-\!\frac{\varphi_2+\varphi'_2}{2\pi}\right)\!\!\left(m_3-\frac{\varphi_3+\varphi'_3}{2\pi}\right).
\end{eqnarray}





If we consider the general case that $f_1\neq 0, f_2\neq 0$ and
$f_3 \neq 0$ with $f_x=f_1+f_2+f_3$, the effective potential can
be obtained straightforwardly as
\begin{eqnarray}
U_{\rm eff}(\{\varphi_i,\varphi'_i\})\!\!\!&=&\!\!\!\frac{3\Phi^2_0}{2L'_{\rm eff}}\left(m_1\!+\!f_1\!-\!\frac{\varphi_1+\varphi'_1}{2\pi}\right)^2\nonumber\\
&+&\frac32\left(\frac{\Phi^2_0}{2L'_{\rm eff}}\!+\!\frac{\Phi^2_0}{2{\tilde L}_{\rm eff}}\right)
\!\!\left[\left( m_2\!+f_2\!-\!\!\frac{\varphi_2+\varphi'_2}{2\pi}\right)^2\!\!+\!\!\left(m_3\!+f_3\!-\!\!\frac{\varphi_3+\varphi'_3}{2\pi}\right)^2   \right]\nonumber\\
&+&\left(\frac{\Phi^2_0}{2L_{\rm eff}}\!-\!\frac{\Phi^2_0}{2L'_{\rm eff}}\right)\left(n\!+\!f_x\!-\!\frac{\varphi_1+\varphi_2+\varphi_3}{2\pi}\right)^2-\sum_i(E_{Ji}\cos\varphi_i+E'_{Ji}\cos\varphi'_i).\nonumber\\
\!&+&\left(\frac{3\Phi^2_0}{2L'_{\rm eff}}\!-\!\frac{3\Phi^2_0}{2{\tilde L}_{\rm eff}}\right)\!\!\left( m_2\!+f_2-\!\frac{\varphi_2+\varphi'_2}{2\pi}\right)\!\!\left(m_3+f_3-\frac{\varphi_3+\varphi'_3}{2\pi}\right).
\end{eqnarray}

\end{document}